# Benchmarking the Nvidia GPU Lineage
From Early K80 to Modern A100 with Asynchronous Memory Transfers

**AUTHOR PREPRINT**

Martin Svedin, Steven W. D. Chien, Gibson Chikafa, Niclas Jansson, Artur Podobas

KTH, Royal Institute of Technology

Stockholm, Sweden

## ABSTRACT

For many, Graphics Processing Units (GPUs) provides a source of reliable computing power. Recently, Nvidia introduced its 9th generation HPC-grade GPUs, the Ampere 100, claiming significant performance improvements over previous generations, particularly for AI-workloads, as well as introducing new architectural features such as asynchronous data movement. But how well does the A100 perform on non-AI benchmarks, and can we expect the A100 to deliver the application improvements we have grown used to with previous GPU generations? In this paper, we benchmark the A100 GPU and compare it to four previous generations of GPUs, with particular focus on empirically quantifying our derived performance expectations, and – should those expectations be undelivered – investigate whether the introduced data-movement features can offset any eventual loss in performance? We find that the A100 delivers less performance increase than previous generations for the well-known Rodinia benchmark suite; we show that some of these performance anomalies can be remedied through clever use of the new data-movement features, which we microbenchmark and demonstrate where (and more importantly, how) they should be used.

## 1 INTRODUCTION

In 2005, Dennard's scaling [3] ended, which forced computer architects to abandon the decade-long reliance on improved clock frequency as the primary source for performance. Instead, the new strategy and immediate remedy was to continue to rely on Moore's law [23], and focus on heterogeneity (specialization) and parallelization (multicore) [2]. Today, this trend has likely reached its climax where programmers and users are bewildered by an ever-increasing amount of heterogeneous accelerators. Device such as Field-Programmable Gate Arrays (FPGAs) are starting to get recognition for their high-performance computing capabilities [5, 20, 22], Coarse-Grained Reconfigurable Architectures (CGRAs) and custom Deep-Learning accelerators are becoming common-place [21], and even alternative computing paradigms such as neuromorphic [24] or quantum systems [11] are emerging. However, among all existing heterogeneos accelerators, none is as ubiquitous as the Graphics Processing Unit (GPU).

Since their introduction in High-Performance Computing (HPC) systems, GPUs have been a steady companion in our road towards exascale computing. Today, seven out of ten of the most powerful machines in the world rely on GPUs to provide the bulk of computing capabilities [1]. The Ampere-100 (A100) [7] is the latest addition to NVidia's Tesla line of high-performance GPUs. Aside from being processed in a smaller process (7nm contra V100's 12nm), the A100 brings several new features to the features, including better virtualization, persistence Static RAM (SRAM) storage across kernel invocations, support for a new numerical representations (a 19-bit format called TensorFloat32), and fine-grained Direct Memory Access (DMA) engine accessible to threads to better overlap computation and communication. Marketing claims often cite up-to 2.7x performance benefits over the previous generation Tesla GPUs, the Volta-100 (V100), often based on applications associated with Deep-Learning (DL) or Artificial Intelligence (AI). But what kind of performance can a user migrating from earlier Tesla-generations expect on the newly released A100– especially if the user is concerned with applications besides AI?

In this paper, we disseminate our experiences in benchmarking the A100 and its new features, particularly honoring the fine-grained asynchronous copy-engines, and quantify the observed gains (or losses) against several previous generations of Tesla-class GPUs. We develop a micro-benchmark for understanding the new asynchronous copy mechanism and apply the knowledge gained from the microbenchmark to four well-known GPU benchmarks.

To our knowledge, this is the first time such a wide benchmark study has been performed on the A100 GPU (the two previous studies focused exclusively on sparse linear solvers [1, 25], and we claim the following contributions:

(1) A quantitative benchmark analysis over several generations of NVidia consumer- and tesla-grade GPUs using the Rodinia benchmark,
(2) A microbenchmark for gauging the performance implications of the new asynchronous copy mechanisms,
(3) Applying and empirically quantifying the impact of using the new asynchronous copy mechanism on four well-known benchmarks using three discretely different approaches

## 2 RELATED WORK

Benchmarking new systems as they become available has a long research lineage, and these works serves as important timestamps to provide computational progressions and position those against existing maxims, common wisdom's, or laws. Despite the recent release of the A100, two benchmarking studies have already been performed. Tsai et al. [25] benchmarked the A100 on sparse linear algebra computation, positioning the performance against the previous V100. Anzt et al. [1] performed a similar study but also included batching, and came to a similar conclusion. Both previous studies showed a 1.8x improvement of A100 over V100.

Previous studies have severely scrutinized predecessors to the A100– something that in time also will happen with the A100. For example, multiple authors have benchmarked the performance performance with [15, 27, 28] or without [16] focus on the (at that time) new tensor units (some even question their applicability, see [9]). Similarly, a newly introduced benchmark suite [14, 19, 26] is often executed with multiple GPUs in hope to capture any new behaviours or insights. Similar to our study, there exists a plethora of microbenchmarking studies trying to capture and isolate a particular details [4], such as memory hierarchies [18], microarchitecture [12, 30] or communication [4].

---
[1] www.top500.org

## 3 OVERVIEW OVER THE A100

The Ampere 100 (A100 from now) is the latest GPU released by Nvidia that targets the HPC, Artificial Intelligence (AI), and Datacenter market. The A100 is part of Nvidia's Tesla family and follows the architectural lineage of its predecessors, including (in reverse chronological order) Nvidia Volta-100, Pascal-100, and Kepler-80. It is built on TSMC's 7 nm technology (same as, e.g., Fujitsu A64FX [31]) and features 54.2 billion transistors spread across 826 $mm^2$ of area. From an architectural perspective, the A100 follows the trend of previous Tesla-generation GPUs, where much focus is placed on parallelism: the A100 has 108 Streaming Microprocessors (called SMs, an increase of 24 SMs over V100), which in turn contain a total of 6912 CUDA cores (nearly double the cores compared to V100); in short, as with previous generations, the architecture is best described as both MIMD [10] and SIMT [2]. The external memory interface is based on HBM2 (1.6 TB/s, ~ 1.7× over V100). While GPUs such as the A100 is primarily focusing on improving the *throughput* of applications (contra a general-purpose CPU which is latency-oriented), the A100 does come with a 40 MB shared L2 cache, which is more than six times as large as the previous V100, reinforcing the line of thought that the gap between CPUs and GPUs is, indeed, shrinking. Among the introduced architectural capabilities are the new elastic GPU features to support multiple GPU instances (that is, virtualization) and better scaling across NVLink. A new numerical representation has been introduced called TensorFlow-32 (TF32), whose name might be slightly misleading since it has little to do with 32-bit, but rather is a custom 19-bit format that has the same exponent width as BFLOAT16 and the same number of mantissa bits as IEEE754 half-precision, and target Deep-Learning. The final architectural improvement – which is also the improvement we empirically evaluate in this paper – is the asynchronous copy mechanism, which aspires to relieve programmers from actively wasting GPU cycles for data-movement, and instead allows the programmer to launch asynchronous data movements that may run in parallel with the computation; essentially, one could view said feature as a improved DMA engine easily accessible from the CUDA program code.

## 4 ASYNCHRONOUS COPY

CUDA 11 and the Ampere architecture introduced support for asynchronous copying of data from global memory to shared memory. The asynchronous memory copies allows the programmer to initiate a transfer of data from global to shared memory, without blocking program execution. Additional primitives are then provided to allow program execution to wait for the asynchronous memory operation to complete. Based on the Nvidia documenta-

---
**Algorithm 1:** Register Bypass

**Data:** Locations of tiles $t_1, ..., t_n$ in global memory

1  temp ← shared array with space for a single tile
2  **for** $i ← 1, ..., n$ **do**
3      initialize asynchronous copy of tile $t_i$ into temp
4      wait for asynchronous copy of $t_i$ to complete
5      synchronize thread block
6      perform computation on temp
7  **end**

---

tion as well as our own attempts at implementing asynchronous copies in various CUDA kernels, we identified three patterns for how one may use asynchronous copies. The first pattern applies to the common situation where one loads data from global memory into a register, only to immediately store it in shared memory. Here we can use asynchronous copies to instead move data directly from global memory to shared memory. In this pattern we do not attempt to use the asynchronous nature of the copies to overlap computation with the memory copies. We refer to this pattern as `Register Bypass`, and pseudocode is provided in Algorithm 1.

---

**Algorithm 2:** Overlap

**Data:** Locations of tiles $t_1, ..., t_n$ in global memory

1  temp ← shared array with space for $k$ tiles
2  **for** $i ← 1, ..., k-1$ **do**
3      initialize async copy of tile $t_i$ into temp[i]
4  **end**
5  **for** $i ← 1, ..., n$ **do**
6      **if** $i + k ≤ n$ **then**
7          initialize async copy of tile $t_{i+k}$ into temp[(i+k) % k]
8      **end**
9      wait for async copy of $t_i$ to complete
10     synchronize thread block
11     perform computation on temp[i % k]
12 **end**

---

**Algorithm 3:** Drop Off

**Data:** Locations of elements $e_1, ..., e_n$ in global memory

1  temp ← shared array with space for blockDim elements
2  initialize async copy of element $e_1$ into temp[threadIdx]
3  **for** $i ← 1, ..., n$ **do**
4      wait for async copy of $e_i$ to complete
5      $e ←$ temp[threadIdx]
6      **if** $i + 1 ≤ n$ **then**
7          initialize async copy of $e_{i+1}$ into temp[threadIdx]
8      **end**
9      perform computation with $e$
10 **end**

---

The next two patterns utilize the asynchronous nature of the copies to overlap computation with the memory transfer, and they differ in when the thread block is synchronized. The first pattern, which we call `Overlap`, initiates a number of asynchronous memory transfers and then waits for the first one to complete. When this occurs, the thread block is synchronized before performing some computation on the data. When the computation completes and if there is more data for the thread block to process, we initiate another

---
[2] A taxonomy coined by Nvidia, which is essentially implicitly managed SIMD.

asynchronous memory transfer. See Algorithm 2 for pseudocode for this pattern.

The last pattern, which we refer to as `Drop off`, differs from `Overlap` in that we do not synchronize the thread block after the asynchronous copy has completed. Instead each thread requests some data to be copied to shared memory, and then waits for this copy to complete. Once the transfer has completed, the thread fetches its data from shared memory into a register and starts an asynchronous copy of the next element to be processed, if there is such an element. Finally the thread performs some computation with the data stored in the register. See Algorithm 3 for pseudocode.

### 4.1 Microbenchmarking Asynchronous Copy

We implemented a synthetic microbenchmark for the `Overlap` pattern described in the previous section. The aim of this microbenchmark is to gain a better understanding of how asynchronous copies compares with using ordinary synchronous memory copies. We wrote two versions of the microbenchmark, one with the Pipeline API and one with Arrive/Wait Barrier that waits for an asynchronous copy to complete.

The utilization of hardware multithreading and switching execution between different warps is the usual answer to how GPUs handle memory latency. To investigate how prefetching with asynchronous copies compares with relying on hardware multithreading, we also consider a second scenario with limited occupancy. We induce low occupancy for the kernels by dynamically allocating as much shared memory as we can, which limits the number of thread blocks that can be concurrently scheduled on an SM.

The microbenchmark operates on 8 GiB of fp32 data. The data is initialized to a random value in the range $[0, 1)$ using cuRAND. The computation that is performed by the microbenchmark consists of element-wise application of the function $x \mapsto \frac{1}{2}x + \frac{1}{2}$ for a variable number of iterations. By changing the number of iterations, the arithmetic intensity of the microbenchmark can be varied.

Our microbenchmark also considers the following additional parameters: size of the tile we read from global to shared memory, number of thread blocks spawned, number of threads per thread block as well as the number of asynchronous copies in-flight simultaneously. When not stated otherwise, we have selected the best possible values for those parameters not mentioned.

Our benchmark runs a large number of CUDA kernels in the same process. We warm up by performing some computations that are not recorded. Between each kernel invocation, we clear the cache by performing computations on a separate set of data.

## 5 EXPERIMENTAL METHODOLOGY

To understand the computation performance of the latest A100 GPU, we perform a benchmark study using selected applications in the Rodinia benchmark suite. We amplified the input set size for most of the benchmark over the default. We positioned the A100's performance against the past four generations of NVIDIA GPU architectures from Kepler to Ampere, for both HPC/data center and consumer versions.

We summarize the characteristics of the chosen GPUs in Table 1[3]. To understand the performance of the GPUs under a wide variety of workloads and access patterns, we use a selected number of applications from the Rodinia benchmark suite [6]. The applications and their characteristics are summarized in Table 2. All the evaluated systems use Linux as the operating system. The host compiler used is GCC and we use either CUDA 8.0, 10.1, or 11.1.

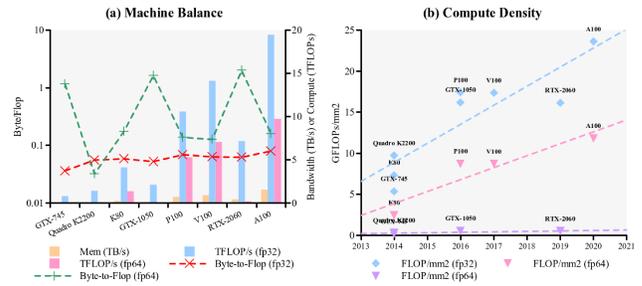

**Figure 1: Derived metrics of evaluated GPUs in terms of: (a) machine balance, and (b) compute densities**

## 6 RESULTS

Before disseminating the performance contrast that A100 experience on the Rodinia benchmarks contra previous GPU generations, it is worthwhile to briefly discuss the trends and differences among them. Figure 1:a overviews the machine balance of the different GPUs, including the amount of memory bandwidth, the compute capacity (fp32 and fp64) and – more importantly – the balance in terms of Byte-to-Flop ratio (B/F) [17]. The B/F ratio is important, as it describes how well an application with a given arithmetic intensity maps to the architecture, and architectures with a low B/F usually means the it is hard to supply data to the (oversized) amount of compute units (see [8] for an example with Intel KNL). Overall, GPUs have a relatively low B/F ratio, often between 0.03- and 0.07 B/F (fp32) and 0.12-0.17 B/F (fp64, contra, e.g., A64FX, which have B/F =0.4), and there is no sign of this changing (for now, Nvidia seem eager to keep this balance), particularly for the Tesla-class GPUs. Consumer-grade GPUs, particularly for fp64 (which are limited in these cards), can have a very high B/F (=2.0 for RTX-2060), making it easy for the programmer to fully use the machine resources (see [13] for an example). Figure 1:b shows the compute density of GPUs in terms of FLOPS/$mm^2$. As expected (in-line with Moore's law), the higher transistor densities also translate to more compute per unit area, which is where Nvidia is spending said transistors. Interestingly, the P100 has a slightly higher compute density than V100. As expected, the amount of fp32 is roughly twice that of fp64 per unit area– this applies to both consumer and Tesla-grade GPUs. Consumer-cards have, however a significantly reduced fp64 capabilities. We end this segment by compiling down our expectations on the A100 GPU. Overall, the increase in compute performance (fp32 and fp64 alike) over the V100 amounts to $FLOP_{v100->a100} = 1.38x$, while the increase in external memory bandwidth amounts to $BW_{v100->a100} = 1.73x$. In conclusion, we can expect at least an improvement of: $T_{speedup} = min(FLOP_{v100->a100}, BW_{v100->a100})$ =**1.38x**, independent of whether the application is memory- or compute-bound (subject to limitations in the benchmarks or toolchains); in the next section, we will see if the A100 delivers the expected improvement.

---
[3]https://www.techpowerup.com/gpu-specs

Table 1: Specifications of the platforms and the GPUs in this evaluation.

| Node Name | Host CPU | Device | Year | Arch. | Interface | Memory | Mem. BW (GB/s) | TFLOP/s (float) | TFLOP/s (double) | # SMs | TDP (Watt) | Die Area ($mm^2$) |
|---|---|---|---|---|---|---|---|---|---|---|---|---|
| Tesla / Data Center GPUs | | | | | | | | | | | | |
| Tegner | Intel Xeon E5-2690v3 | K80 (one GK210) | 2014 Q4 | Kepler | PCIe 3.0 | 12 GB GDDR5 | 240.6 | 4.113 | 1.371 | 13 | 300 | 561 |
| P100 | IBM POWER8 | P100-SMX2 | 2016 Q2 | Pascal | PCIe 3.0 | 16 GB HBM2 | 732.2 | 10.61 | 5.304 | 56 | 300 | 610 |
| Kebnekaise | Intel Xeon Gold 6132 | V100 | 2017 Q3 | Volta | PCIe 3.0 | 16 GB HBM2 | 897.0 | 14.13 | 7.066 | 80 | 300 | 815 |
| A100 | AMD EPYC 7302P | A100 | 2020 Q3 | Ampere | PCIe 4.0 | 40 GB HBM2 | 1555 | 19.49 | 9.746 | 108 | 250 | 826 |
| Workstation / Consumer GPUs | | | | | | | | | | | | |
| GTX745 | Intel i7-6700 | GTX 745 (OEM) | 2014 Q1 | Maxwell | PCIe 3.0 | 4 GB DDR3 | 28.80 | 0.793 | 0.02479 | 3 | 55 | 148 |
| K2200 | Intel Xeon E5-1620v3 | Qaudro K2200 | 2014 Q3 | Maxwell | PCIe 2.0 | 4 GB GDDR5 | 80.19 | 1.439 | 0.04496 | 5 | 68 | 148 |
| Blackdog | Intel Xeon E5-2609v2 | GTX 1050 Ti | 2016 Q4 | Pascal | PCIe 3.0 | 4 GB GDDR5 | 112.1 | 2.138 | 0.0668 | 6 | 75 | 132 |
| Greendog | Intel Core i7-7820X | RTX 2060 SUPER | 2019 Q3 | Turing | PCIe 3.0 | 8 GB GDDR6 | 448.0 | 7.181 | 0.224 | 34 | 175 | 445 |

Table 2: Overview of the selected benchmarks from the Rodinia benchmark suite. We Report here the kernels that are timed and their input parameters.

| Name | Type | Kernels/functions | Input |
|---|---|---|---|
| Backprop | ML | bpnn_adjust_weights_cuda(), bpnn_layerforward_CUDA() | $2^{20} - 16$ |
| BFS | Graph | kernel(), kernel2() | graph16M |
| CFD | Solver | cudaMemcpy(D2D), cuda_compute_step_factor(), cuda_compute_flux(), cuda_time_step() | fvcorr.domn.193K |
| Gaussian | Solver | Fan1(), Fan2() | 8192 |
| Heartwall | Imaging | kernel() | test.avi, 104 |
| Hotspot | Physics | calculate_temp() | $8192^2$, pyramid=2, iter=5 |
| Hotspot3D | Physics | hotspotOpt1() | $4096^2 \times 8$, iter=10 |
| lavaMD | Simulation | kernel_gpu_cuda() | boxes1d=80 |
| Leukocyte | Imaging | IMGVF_kernel() | testfile.avi, 599 |
| LUD | Solver | lud_diagonal(), lud_perimeter(), lud_internal(), lud_diagonal() | matrix_size=16384 |
| Myocyte | Simulation | kernel() | xmax=100, workload=10, mode=0 |
| NW | Optimization | needle_cuda_shared_1(), needle_cuda_shared_2() | row/col=16384, penalty=10 |
| PathFinder | Optimization | dynproc_kernel() | row=100000, col=10000, pyramid=20 |
| Srad (v1) | Imaging | prepare(), reduce(), cudaMemcpy(D2H), srad(), srad2() | iter=1000, lambda=0.5, Nr=2160, Nc=3840 |
| StreamCluster | Data Mining | kernel_compute_cost() | k1=10, k2=20, d=256, n=65536, chunksize=65536, clustersize=1000, infile=none, outfile=output.txt, nproc=1 |

## 6.1 Rodinia Performance

Figure 2 shows the quantitative evaluation of Rodinia when executed with multiple different GPUs. Consumer GPUs are clustered to the left in graphs, while HPC/Data-center GPUs to the right. In total, we have four GPUs representing each category. Overall, we see an expected trend of improvement when moving across generations. The execution of the benchmark is fairly stable, with the exception of the early GTX-745 on Backpropagation and BFS, whose standard deviation is on the order of 2%. The kernels' execution time varied from an order of milliseconds (e.g., Backpropagation) to several seconds (e.g., CFD); the low execution time of some kernels – despite our efforts to enlarge the problem size of the benchmarks – reveals an alarming direction of using Rodinia for future benchmarking of GPUs.

The largest average improvement is when upgrading from a K80→P100, which gives a ∼ 3.95× speed-up, tightly followed by going from GTX 1050-Ti→RTX 2060, which yields ∼ 3.36× speed-up. For the earlier, this increase is likely because we are moving two generations (leap-frogging Maxwell, which was not available to us), while the latter is due to the ∼ 4× larger die size (the compute density of GTX 1050 vs. RTX 2060 is actually roughly the same). Interestingly, the A100 experience on average 1.34x improvement– the lowest improvement among our observed historical upgrades and lower than the minimum expected improvement of 1.38x. The A100 performs particularly suboptimal on the Heartwall, HotSpot, Leukocyte, Myocyte, and Pathfinder applications. The best performing applications with the A100 are BFS (∼ 2.76× over V100), CFD (∼ 1.89× over V100), and SRAD (∼ 1.5× over V100). A condensed view over the upgrade improvements of the cards is seen in Figure 2, where we highlight our minimal expected performance increase of moving from V100→A100– as can be seen, overall, the A100 does perform better than V100 but does not deliver as much of an improvement when compared with previous generations.

Another interesting observation is that the gap in performance – at least on the Rodinia benchmarks – between consumer-grade and HPC-grade generations is indeed shrinking. For example, the average speed-up of K80 (Kepler) over GTX 745 (Maxwell) is 4.28x ± 1.83x. However, the average speed-up of V100 (Volta) or A100 (Ampere) over RTX 2060 (Turing) is 2x± 0.49x and 2.66x±0.97x respectively, which is a significant decrease in performance difference even with one generation ahead.

The underlying reason behind said performance discrepancies (of the A100) are hard to concretize without further scrutiny. Among the potential culprits could be the (yet-to-mature) toolchain and compiler, the Rodinia benchmarks, or simply the fact that the manufacturer chose to focus on the other aspects (e.g., Deep-Learning) in the architecture. At the same time, perhaps the newly introduced data-movement options can offset the loss in performance?

## 6.2 Asynchronous Copy Performance

Can the asynchronous copy improve performance, and – if so – when? We configured our microbenchmark to contain various degrees of arithmetic intensity ($\frac{FLOPS}{Byte}$), and executed the benchmark without any asynchronous copies (standard), using barriers (see section 4), and using pipelines. The results, positioned against the roofline [29], are seen in Figure 3:a. Overall, we see that when the microbenchmark is configured with low arithmetic intensity, using asynchronous copy can help the performance to reach closer to the expected roofline, yielding 1.07x-1.35x (avg: 1.27x) and 1.1x-1.4x (avg: 1.33x) for barriers and pipeline respectively. At higher arithmetic intensity, the asynchronous copy merely adds overhead to already compute-bound microbenchmark, and benefits are no longer observed; instead, performance is slightly degraded (no-more than 0.95x). In conclusion, the microbenchmark shows that asynchronous copies *can*, to some extent, remedy performance problems bound by memory by pushing the performance closer to the roof.

Figure 3:b-c shows the impact on speed-up (relative to not using async) as a function of per-CUDA-block working size. We observe that the general trend is that when the arithmetic intensity is low, and the data-footprint is sufficiently high, both the barrier and pipeline versions yield speed-ups, but when the microbenchmark becomes compute-bound, we see a performance degradation (occurring according to roofline, see Figure 3:a). Using pipeline is observed to be better and more stable barrier, with a both high speedup for low arithmetic intensity and a more graceful degradation when compute-bound, encouraging users to use the pipeline

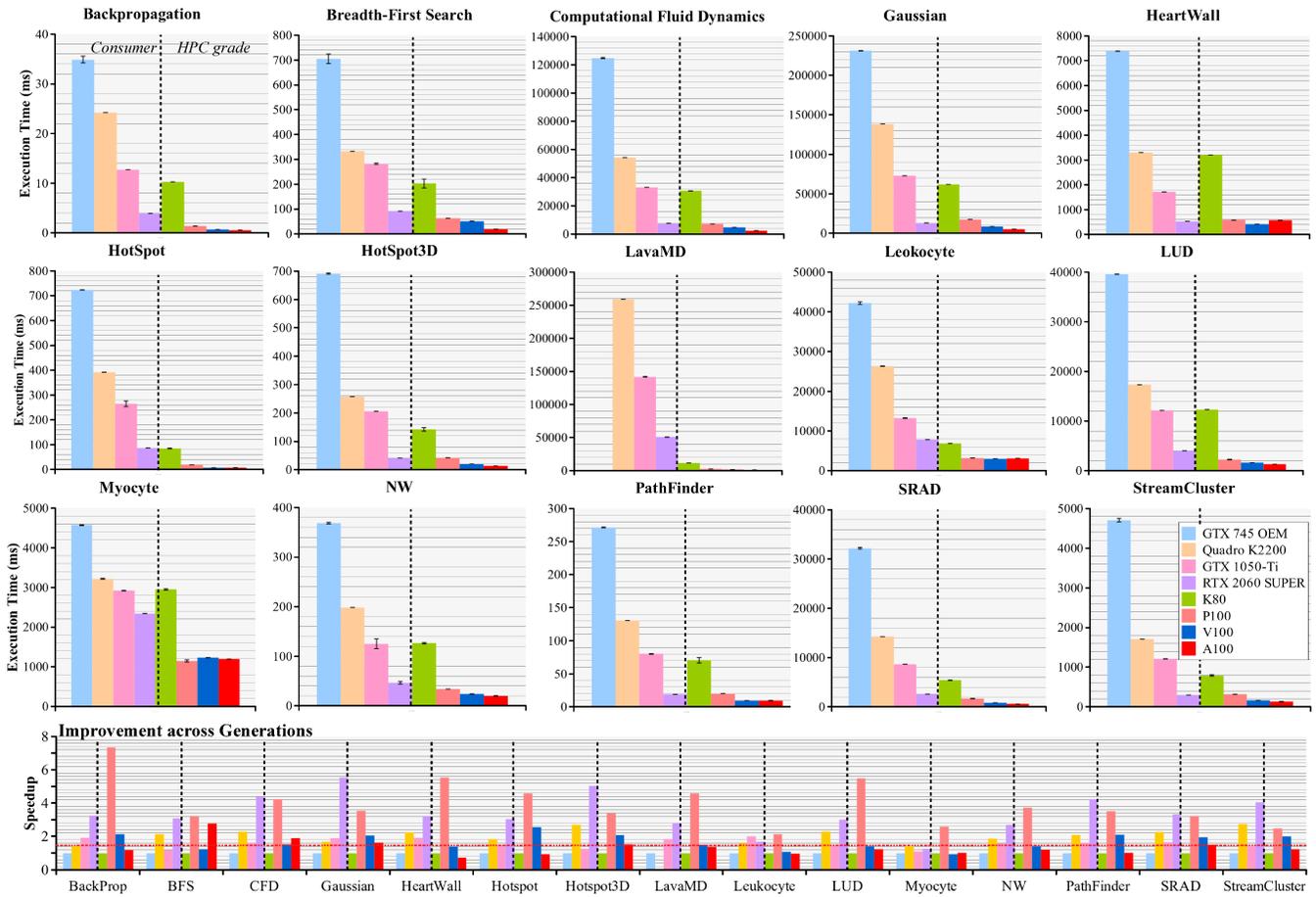

Figure 2: Performance of individual Rodinia benchmarks with six different GPUs spanning four generations (top), and the average speed-up we experience when moving from one generation to another (bottom, $GTX745 \to RTX2060$, and $K80 \to A100$).

over barrier. barrier requires quite large working-sets (> 256) in order to yield any performance increases.

Figure 3:d shows how the execution when we try to capture the use of asynchronous copies in *low occupancy* scenarios– scenarios where fewer threads have to do more work, including fetching data from memory. We observe that in low occupancy scenarios, not using asynchronous copy can nearly double the execution time (half the performance) of the microbenchmark; this is because threads spend their both fetching data from memory and computing on it, which can be particularly hurtful at low arithmetic intensities. On the other hand, by cleverly outsourcing the data movement to the asynchronous memory transfers, even with low occupancy, we can limit this performance penalty to between 0.7-17.19% and 2.28-26.56% for barrier and pipeline respectively. Note that pipeline is still executing faster than barrier, but does seem slightly (avg: 14.24%) less resilience to low occupancy than barrier (avg: 6.85%).

Can the performance we observe in our microbenchmark also materialize in the Rodinia benchmarks?

### 6.3 Extending Rodinia with Async. Copies

We applied our strategies in Section 4.1 to four Rodinia benchmarks: NW, LUD, Hotspot, and Pathfinder. Note that not all of our strategies can be applied to each benchmark. All four chosen benchmarks experienced lower-than-expected performance comparing V100 to A100 (see previous section). The observed improvement of applying the asynchronous copies can be seen in Figure 4. Overall, we can observe that at least one (out of three) of our strategies can improve the performance of the benchmarks; the improvement and strategy seem subject to both which benchmark it is and the input size. Consider, for example, Hotspot, which benefits primarily from applying the Overlap strategy, improving performance by 1.12x-1.23x over baseline. In NW, the opposite holds true, and Register Bypass is the strategy of choice, yielding between 1.01x-1.08x improvements over baseline. The LUD benchmark is interesting, as the strategy of choice is dependent on the input size: for 8192x8192 matrices, the Register Bypass yield an improvement of 1.32x over baseline; as we increase the input size, we noticed that Overlap starts to become more effective (yielding 1.25-1.27x speedup over baseline), overtaking Register Bypass. This shows that the strategies can be effective at different working size. The Pathfinder benchmark was the only (out of the four) amendable to the `Drop Off` strategy, which yields between 1.04x-1.11x improvement (and, for some reason, suffers from rather large standard deviations, subject to future scrutiny). Finally, we end by saying that our results point towards further research opportunities in better understanding exactly where, how, and when asynchronous data movement should be used.

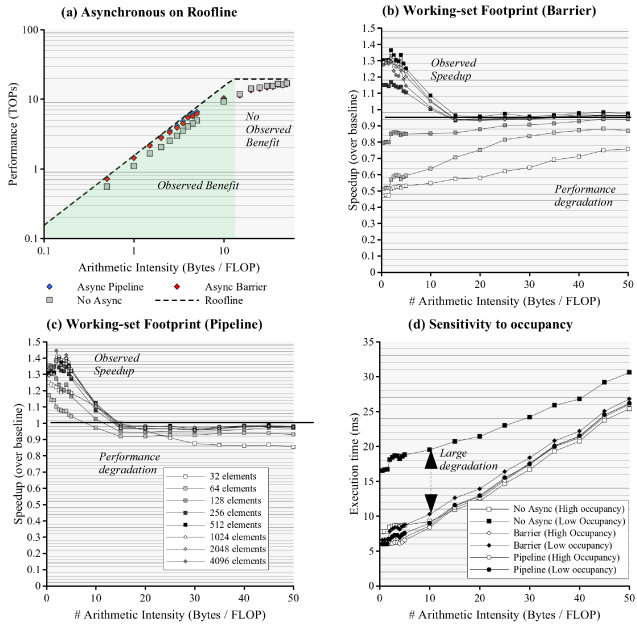

Figure 3: The impact on execution time of integrating various asynchronous memory strategies (Section 4.1) on four Rodinia benchmarks.

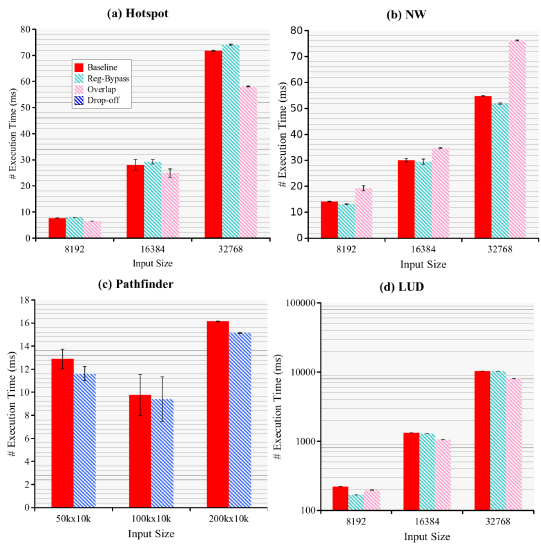

Figure 4: The impact on execution time of integrating various asynchronous memory strategies (Section 4.1) on four Rodinia benchmarks.

## 7 CONCLUSION

In this paper, we have studied the performance of the recent Nvidia A100 on a diverse set of benchmarks. We empirically positioned and contrasted the performance of A100 against earlier generations in the Nvidia GPU lineage and found that A100 – on the current set of benchmarks – is slightly underperforming compared to expectations. We further microbenchmark the new asynchronous data movement support and applied obtained knowledge to a subset of the benchmarks, showing that the new data movement construct has two edges and can improve performance (we observed up to 1.25x) but can also harm it.